\documentclass[graybox]{svmult}

\usepackage{mathptmx}       % selects Times Roman as basic font
\usepackage{helvet}         % selects Helvetica as sans-serif font
\usepackage{courier}        % selects Courier as typewriter font
\usepackage{makeidx}         % allows index generation
\usepackage{graphicx}        % standard LaTeX graphics tool
\usepackage{multicol}        % used for the two-column index
\usepackage[bottom]{footmisc}% places footnotes at page bottom

\makeindex             % used for the subject index
\begin{document}

\title*{Investigation of Rotational Splitting in the Pulsating White Dwarf GD 154}
\author{Zs. Bogn\'ar and M. Papar\'o}
\institute{Zs\'ofia Bogn\'ar \at Konkoly Observatory of the Hungarian Academy of Sciences, H-1525 Budapest, P.O. Box 67., \email{bognar@konkoly.hu}}
%\and Margit Papar\'o \at Konkoly Observatory of the Hungarian Academy of Sciences, H-1525 Budapest, P.O. Box 67., \email{paparo@konkoly.hu}}
\maketitle
% Please use both starred abstract and non-starred abstract.

\abstract*{We observed the ZZ Ceti star GD 154 over a whole season at the mountain station of Konkoly Observatory. Our long time base allowed to detect the sign of rotational triplets around the independent modes. To check whether these can be real detections we made a test on our data set. We searched for characteristic spacing values performing Fourier analysis of numerous peaks determined around five frequencies in the main pulsation region. The analysis revealed regular peak spacings with separations around 3.7 and 2.6\,$\mu$Hz. These values are in accordance with the ones determined by consecutive prewhitening of Whole Earth Telescope observations.}

\abstract{We observed the ZZ Ceti star GD 154 over a whole season at the mountain station of Konkoly Observatory. Our long time base allowed to detect the sign of rotational triplets around the independent modes. To check whether these can be real detections we made a test on our data set. We searched for characteristic spacing values performing Fourier analysis of numerous peaks determined around five frequencies in the main pulsation region. The analysis revealed regular peak spacings with separations around 3.7 and 2.6\,$\mu$Hz. These values are in accordance with the ones determined by consecutive prewhitening of Whole Earth Telescope observations.}

\section{The Method and the Test Object}
\label{sec:1}

Handler et al. (1997) presented a method to investigate the regularities in the distribution of the pulsation frequencies of a $\delta$ Scuti star. They assigned unit amplitude to all of the frequencies and calculated their Fourier transform (FT). In the resulting window function, they detected a characteristic spacing and determined its value. This method was applied later in the case of other stars, too. These results gave us the inspiration to test how this method is applicable for rotationally split frequencies.

GD 154 has been chosen as our test object. It's brightness variations were discovered in 1977 (Robinson et al. 1978). In 1991, GD 154 was the target of a Whole Earth Telescope campaign. Using this dataset Pfeiffer et al. (1996) determined three independent modes and they have also found frequencies supposed to be rotational split components. The frequency separations at the dominant peak were 2.8 and 2.2\,$\mu$Hz, respectively. They have also found a peak near to the second largest amplitude mode with a separation of 3.9\,$\mu$Hz. Using a two-site campaign's data set, H\"urkal et al. (2005) have also found the signs of rotational splitting at two modes with 3.1, 3.4 and 3.3\,$\mu$Hz frequency separations, respectively.

\section{New Observations and Our Results}
\label{sec:2}

We observed GD 154 on a time base of six months at Piszk\'estet\H o mountain station of Konkoly Observatory. We detected the sign of rotational triplets around different modes. To check whether we found real triplet components, we performed a test and searched for the characteristic frequency splitting value(s) adopting Handler's method.

We pre-whitened the Fourier spectrum of the whole dataset with five significant frequencies in the main pulsation region. Then we searched for peaks in the residual spectrum in the $\pm$10\,$\mu$Hz vicinity of these modes, and determined the two largest amplitude ones in every single frequency range applying the usual successive pre-whitening method. The frequency list obtained finally consisted of 3$\times$5 elements -- the five pulsation modes and the two additional frequencies per domains. We assigned unit amplitude to all of the frequencies and generated their FT. To check the stability of our results, we performed the described analysis by the omission of one or two frequency domains, too.

Using all of the 15 peaks, a frequency at 3.7\,$\mu$Hz and its harmonic have the largest amplitudes in their FT. In the case of the three- or four-domain solutions the peak at 3.7\,$\mu$Hz remains detectable and additional ones appear with relatively high amplitudes at 2.5--2.7\,$\mu$Hz and 3.9--4.1\,$\mu$Hz.

We checked the light curve's window function and it does not show any high amplitude peaks around 3.7--4.1 or 2.5--2.7\,$\mu$Hz, comparing to the main peaks around zero or the 1-day alias. This means that technical peaks do not disturb our spacing determinations.

Based on this investigation, characteristic spacings with values around 2.6 and 3.7\,$\mu$Hz may present in our data set. These values are in accordance with the frequency separations determined by previous observations and considered as the results of rotational splitting. This suggests that with this indirect method we may be able to detect the sign of frequency splittings caused by the rotation of a star. 

%\begin{figure}[b]
%\sidecaption
% For example, with the graphicx style use
%\includegraphics[scale=.4]{abra_1.eps}
%
% If no graphics program available, insert a blank space i.e. use
%\picplace{5cm}{2cm} % Give the correct figure height and width in cm
%
%\caption{zzz}
%\end{figure}

% 

\end{document}